\input harvmac


\def\N{{\cal N} }

\def \lr {\lref }

\def \pw {plane-wave\ }
\def \eq#1 {\eqno{(#1)}}

\def \a {\alpha}

\def \om {\omega}

\def \s {\sigma}

\def \e#1 {{\rm e}^{#1}}

\def \ha { { 1\over 2 }}

\def \td {\tilde}

\def\p{\partial }
\def\four{ {\textstyle{1\over 4}}}
\def\K{ {\cal K} }


\baselineskip8pt \Title{\vbox {\baselineskip 6pt{\hbox{ }}{\hbox {
}}{\hbox {}}{\hbox{ }} {\hbox{ }}} } {\vbox{\centerline { Anomalous dimensions in gauge theories} \vskip4pt \centerline{  from  rotating strings in $AdS_5\times S^5$
} }} \vskip -20 true pt
\centerline { {J.G. Russo \footnote {$^*$} {e-mail address:
russo@df.uba.ar  } }}

\medskip \smallskip
\centerline {\it {} Departamento de F\'\i sica, Universidad de
Buenos Aires, }
\smallskip
\centerline {\it Ciudad Universitaria and Conicet, Pab. I, 1428
Buenos Aires, Argentina}

\bigskip
\centerline {\bf Abstract}
\medskip
\baselineskip10pt \noindent
Semi-classical soliton solutions  for superstrings in $AdS_5\times S^5$ are used
to predict the dimension of gauge theory operators in $\N=4 $
SU(N) SYM theory. We discuss the possible origin of scaling
violations
on the gauge theory side.

\medskip

\Date {May 2002}
\noblackbox
\baselineskip 16pt plus 2pt minus 2pt


\noblackbox \overfullrule=0pt

\def \rf {\refs}

\def\four{{\textstyle{1\over 4} }}



\lref\hrts{G.~T.~Horowitz and A.~A.~Tseytlin, ``A New class of
exact solutions in string theory,'' Phys.\ Rev.\ D {\bf 51}, 2896
(1995) [hep-th/9409021].
}

\lref\russu{ J.~G.~Russo and L.~Susskind, ``Asymptotic level
density in heterotic string theory and rotating black holes,''
Nucl.\ Phys.\ B {\bf 437}, 611 (1995) [hep-th/9405117].
}

\lref\polya{ A.~M.~Polyakov,
 ``Gauge fields and space-time,'' arXiv:hep-th/0110196.
}

\lref\parna{A.~Parnachev and D.~A.~Sahakyan,
``Penrose limit and string quantization in AdS(3) x S**3,''
arXiv:hep-th/0205015.
}

\lref\lee{
P.~Lee and J.~W.~Park,
``Open strings in PP-wave background from defect conformal field theory,''
arXiv:hep-th/0203257.
}

\lref\minw{
N.~R.~Constable, D.~Z.~Freedman, M.~Headrick, S.~Minwalla, L.~Motl, A.~Postnikov and W.~Skiba,
``PP-wave string interactions from perturbative Yang-Mills theory,''
arXiv:hep-th/0205089.
}

\lref \napwi { C.~R.~Nappi and E.~Witten, ``A WZW model based on a
nonsemisimple group,'' Phys.\ Rev.\ Lett.\ {\bf 71}, 3751 (1993)
[hep-th/9310112].
}

\lref \rrt { J.~G.~Russo and A.~A.~Tseytlin, ``Magnetic flux tube
models in superstring theory,'' Nucl.\ Phys.\ B {\bf 461}, 131
(1996) [hep-th/9508068].
}

\lref \tserus{ J.~G.~Russo and A.~A.~Tseytlin, ``Constant magnetic
field in closed string theory: An Exactly solvable model,'' Nucl.\
Phys.\ B {\bf 448}, 293 (1995) [hep-th/9411099].
}

\lref\russo{J.~G.~Russo,
``The string spectrum on the horizon of a non-extremal black hole,'' Nucl.\ Phys.\ B {\bf 481}, 743 (1996)
[arXiv:hep-th/9606031].
}

\lr\tseyt{A.A.~Tseytlin, Nucl.\ Phys.\ B {\bf 477}, 118 (1996).}

\lr\TT{ A.~A.~Tseytlin, ``Exact solutions of closed string
theory,'' Class.\ Quant.\ Grav.\ {\bf 12}, 2365 (1995)
[hep-th/9505052].
}

\lr\super{J.~G.~Russo and A.~A.~Tseytlin, ``Supersymmetric
fluxbrane intersections and closed string tachyons,'' JHEP {\bf
0111}, 065 (2001) [hep-th/0110107].}

\lr\ruti{J.~G.~Russo and A.~A.~Tseytlin,
``On solvable models of type IIB superstring in NS-NS and R-R plane wave backgrounds,''
arXiv:hep-th/0202179.
}

\lr\sdas{S.~R.~Das, C.~Gomez and S.~J.~Rey,
 ``Penrose limit, spontaneous symmetry breaking and holography in pp-wave background,'' arXiv:hep-th/0203164.
}

\lr\leigh{ R.~G.~Leigh, K.~Okuyama and M.~Rozali,
``PP-waves and holography,'' arXiv:hep-th/0204026.
}

\lr\green{ O.~Bergman, M.~R.~Gaberdiel and M.~B.~Green,
``D-brane
interactions in type IIB plane-wave background,''
arXiv:hep-th/0205183.
}

\lr \kkl {E.~Kiritsis, C.~Kounnas and D.~Lust, ``Superstring
gravitational wave backgrounds with space-time supersymmetry,''
Phys.\ Lett.\ B {\bf 331}, 321 (1994) [hep-th/9404114].
}

\lref \rshet{ J.~G.~Russo and A.~A.~Tseytlin, ``Heterotic strings
in uniform magnetic field,'' Nucl.\ Phys.\ B {\bf 454}, 164 (1995)
[hep-th/9506071].}

\lref\devega{H.~J.~de Vega and I.~L.~Egusquiza,
``Planetoid String Solutions in 3 + 1 Axisymmetric Spacetimes,'' Phys.\ Rev.\ D {\bf 54}, 7513 (1996) [arXiv:hep-th/9607056].
}

\lref\suga {Y. Hikida and Y. Sugawara,
``Superstrings on PP-Wave Backgrounds and Symmetric Orbifolds,'' [hep-th/0205200].
}

\lref \gsw { M.~B.~Green, J.~H.~Schwarz and E.~Witten,
``Superstring Theory. Vol. 1: Introduction,'' {\it Cambridge, UK:
Univ. Pr. ( 1987)}, section 2.3.5. }

\lref\rrm { R.~R.~Metsaev, ``Type IIB Green Schwarz superstring in
plane wave Ramond Ramond background,'' hep-th/0112044.
}

\lref \lust { E.~Kiritsis, C.~Kounnas and D.~Lust, ``Superstring
gravitational wave backgrounds with space-time supersymmetry,''
Phys.\ Lett.\ B {\bf 331}, 321 (1994) [hep-th/9404114].
}

\lref \blau { M.~Blau, J.~Figueroa-O'Farrill, C.~Hull and
G.~Papadopoulos, ``A new maximally supersymmetric background of
IIB superstring theory,'' hep-th/0110242.
}

\lr\blap{ M.~Blau, J.~Figueroa-O'Farrill, C.~Hull and
G.~Papadopoulos, ``Penrose limits and maximal supersymmetry,''
hep-th/0201081.
}

\lr\gross{D.~J.~Gross, A.~Mikhailov and R.~Roiban,
``Operators with large R charge in N = 4 Yang-Mills theory,''
arXiv:hep-th/0205066.
}

\lr \mald { D.~Berenstein, J.~Maldacena and H.~Nastase, ``Strings
in flat space and pp waves from N = 4 super Yang Mills,''
hep-th/0202021. }

\lr\itza{N.~Itzhaki, J.~M.~Maldacena, J.~Sonnenschein and S.~Yankielowicz,
``Supergravity and the large N limit of theories with sixteen supercharges,'' Phys.\ Rev.\ D {\bf 58}, 046004 (1998) [arXiv:hep-th/9802042].
}

\lr \hull {C.~M.~Hull, ``Killing Spinors And Exact Plane Wave
Solutions Of Extended Supergravity,'' Phys.\ Rev.\ D {\bf 30}, 334
(1984).
J.~Kowalski-Glikman, ``Vacuum States In Supersymmetric
Kaluza-Klein Theory,'' Phys.\ Lett.\ B {\bf 134}, 194 (1984).
R.~Gueven,
``Plane Waves In Effective Field Theories Of Superstrings,''
Phys.\ Lett.\ B {\bf 191}, 275 (1987).
}

\lr \mt { R.~R.~Metsaev and A.~A.~Tseytlin, ``Exactly solvable
model of superstring in plane wave Ramond-Ramond background,''
hep-th/0202109.
}

\lr\others{Other recent works on RR pp waves...}


\lr\papa{ M.~Blau, J.~Figueroa-O'Farrill and G.~Papadopoulos,
``Penrose limits, supergravity and brane dynamics,''
hep-th/0202111.
}

\lr \itz{N. Itzhaki, I.R. Klebanov and S. Mukhi, ``PP Wave Limit
and Enhanced Supersymmetry in Gauge Theories'', hep-th/0202153.}

\lr \oog{ J. Gomis and H. Ooguri, ``Penrose Limit of N=1 Gauge
Theories'', hep-th/0202157.}

\lr\guv{R.~Gueven, ``Plane wave limits and T-duality,'' Phys.\
Lett.\ B {\bf 482}, 255 (2000) [hep-th/0005061].
}

\lref \mplt { A.~A.~Tseytlin, ``Extreme dyonic black holes in
string theory,'' Mod.\ Phys.\ Lett.\ A {\bf 11}, 689 (1996)
[hep-th/9601177].
}

\lref\pando{L.~A.~Zayas and J.~Sonnenschein, ``On Penrose limits
and gauge theories,'' JHEP {\bf 0205}, 010 (2002)
[arXiv:hep-th/0202186].
}

\lref\alish{M.~Alishahiha and M.~M.~Sheikh-Jabbari,
``The PP-wave limits of orbifolded AdS(5) x S**5,''
arXiv:hep-th/0203018.
}

\lref\kim{N.~w.~Kim, A.~Pankiewicz, S.~J.~Rey and S.~Theisen,
``Superstring on pp-wave orbifold from large-N quiver gauge
theory,''
arXiv:hep-th/0203080.
 }

\lref\taka{T.~Takayanagi and S.~Terashima,
``Strings on orbifolded pp-waves,'' arXiv:hep-th/0203093.
}

\lref\flo{E.~Floratos and A.~Kehagias,
``Penrose limits of
orbifolds and orientifolds,''
arXiv:hep-th/0203134.
}

\lref\gava{ D.~Berenstein, E.~Gava, J.~M.~Maldacena, K.~S.~Narain
and H.~Nastase,
``Open strings on plane waves and their Yang-Mills duals,''
arXiv:hep-th/0203249.
}

\lref\mich{J.~Michelson,
``(Twisted) toroidal compactification of pp-waves,'' hep-th/0203140.
}

\lref\imamura{Y.~Imamura, ``Large angular momentum closed strings
colliding with D-branes,''
arXiv:hep-th/0204200.
}

\lr\oogm{ J.~Maldacena and H.~Ooguri,
``Strings in AdS(3) and SL(2,R) WZW model. I,''
J.\ Math.\ Phys.\ {\bf 42}, 2929 (2001)
[hep-th/0001053].
``Strings in AdS(3) and the SL(2,R) WZW model. III: Correlation functions,''
hep-th/0111180.
}

\lr\maldacena{J.~Maldacena,
``The large $N$ limit of superconformal field theories and supergravity,''
Adv.\ Theor.\ Math.\ Phys.\ {\bf 2}, 231 (1998) [Int.\ J.\ Theor.\ Phys.\ {\bf 38}, 1113 (1998)]
[arXiv:hep-th/9711200].
}

\lr \gkp {S.~S.~Gubser, I.~R.~Klebanov and A.~M.~Polyakov,
``Gauge theory correlators from non-critical string theory,''
Phys.\ Lett.\ B {\bf 428}, 105 (1998) [arXiv:hep-th/9802109].
}

\lr\witten{E.~Witten,
``Anti-de Sitter space and holography,''
Adv.\ Theor.\ Math.\ Phys.\ {\bf 2}, 253 (1998) [arXiv:hep-th/9802150].
}

\lr\witt{E.~Witten,
``Anti-de Sitter space, thermal phase transition, and confinement in gauge theories,''
Adv.\ Theor.\ Math.\ Phys.\ {\bf 2}, 505 (1998) [arXiv:hep-th/9803131].
}

\lr\newrusso{J.~G.~Russo,
``New compactifications of supergravities and large N {QCD},'' Nucl.\ Phys.\ B {\bf 543}, 183 (1999)
[arXiv:hep-th/9808117].
}

\lr\rotating{ C.~Csaki, Y.~Oz, J.~Russo and J.~Terning,
``Large N {QCD} from rotating branes,'' Phys.\ Rev.\ D {\bf 59}, 065012 (1999) [arXiv:hep-th/9810186].
}

\lr\gkpsol{S.~S.~Gubser, I.~R.~Klebanov and A.~M.~Polyakov,
``A semi-classical limit of the gauge/string correspondence,'' arXiv:hep-th/0204051.
}

\lr\ft {S.~Frolov and A.~A.~Tseytlin,
``Semiclassical quantization of rotating superstring in AdS(5) x S**5,''
arXiv:hep-th/0204226.
}


\newsec{Introduction}

Understanding  the AdS/CFT correspondence \refs{\maldacena, \gkp
, \witten} to full string theory level can lead to important
insights about the $1/N$ expansion and strong coupling physics in
Yang-Mills theories. It is therefore of interest to develop
methods to study  string propagation in  backgrounds with R-R
gauge fields. Recently, some direct checks of the AdS/CFT
correspondence beyond the supergravity level were performed by
considering a special sector of states where some global quantum
numbers (such as R-charge or spin) are large \refs{\mald ,\polya,
\gkpsol , \gross }.
The idea of \mald\ is based on the observation
\refs{\rrm,\mald,\mt,\ruti} that string theory  in
certain \rf{\blau,\blap} R-R \pw backgrounds is solvable.
Extensions in different directions
were carried out in
\refs{\itz\oog\pando\alish\kim\taka\flo\mich\sdas\gava\lee\leigh\imamura\parna\minw\green -\suga }.

The strategy of \gkpsol , which was further elaborated in \ft , is
to identify certain semi-classical soliton solutions representing highly
excited string states with gauge theory operators of
some finite anomalous dimension.
The classical energy in global $AdS$ coordinates is identified with the
conformal dimension of the corresponding state in the dual gauge theory.

The soliton solutions investigated in \gkpsol\
can be separated into three classes:
\smallskip

\noindent i) Strings spinning on $AdS_5$,  stretched along the
radial direction $\rho $, at fixed angles on $S^5$ (an earlier
study of this classical string solution is in \devega ). They
represent string states
with spin $S$, and they have an energy
\eqn\runo{
E\cong S+{\sqrt{\lambda }\over\pi }
\log {S\over\sqrt{\lambda }} \ ,\ \ \ \ S\gg \sqrt{\lambda }\ .
}

\noindent ii) Strings spinning
on $S^5$, stretched along the radial direction $\rho $,
representing string states carrying R-charge $J$. For them
 \eqn\rdos{ E=J
+\sum_{n=-\infty}^\infty  N_n \sqrt{1+{\lambda n^2\over J^2} }\ ,\ \ \ \ J\gg \sqrt{\lambda }\  ,
}
where the second term arises by considering small oscillations
around the soliton state. This is the same formula obtained by
\mald\ in the exactly solvable \refs{\rrm , \mt } pp-wave background
\refs{\blau,\blap}.

\noindent iii) Strings spinning on $S^5$ (i.e. with R-charge
$J$), sitting at $ \rho =0$ and
 strectched along an angular direction.
Here one obtains \gkpsol \eqn\rtres{ E\cong J+{2\sqrt{\lambda}
\over\pi }\ . }
\smallskip
\noindent
In \ft , a more general solution interpolating between the cases i) and ii) was investigated, leading to a general formula for the energy as a function of $S$ and $J$.
Here we shall present a solution interpolating between the three cases i), ii), iii) described above.
This will lead to a more general formula $E=E(S,J)$, which will reduce to the previously known cases by taking different limits.
The  string state (and thus the corresponding gauge operator)  is not the same as the one studied in \ft .
In a large $J$ limit, the energy approaches the energy of the state of \ft .
The main virtue of the present approach is to incorporate the features of the formulas \runo -\rtres\ into a single general expression for the conformal dimension.

\newsec{More general soliton rotating in $AdS_5$ and $S^5$}

We consider type IIB superstrings moving in the $AdS_5\times S^5$ background, with metric:
\eqn\metric{
ds^2=\a' R^2 \big[ -dt^2 \cosh^2\rho +d\rho^2+\sinh^2\rho d\td \Omega_3^2
+d\psi^2 \sin^2\theta +d\theta^2+\cos^2\theta d \Omega_3^2\big]\ ,
}
$$
d\td \Omega_3^2=\cos^2\beta d\td \phi ^2+ d\beta^2 +\sin^2\beta d\phi^2\ ,\
$$
$$
d \Omega_3^2=\cos^2\psi_1 d\psi_2 ^2+ d\psi_1^2 +\sin^2\psi_1 d\psi_3^2\ ,\
$$
where $R^2=\sqrt{\lambda }$ and $\lambda =g^2_{\rm YM}N$ is the 't Hooft coupling.
We look for a soliton solution of the following form:
\eqn\anzz{
t=\kappa \tau\ ,\ \ \ \phi=\om \tau\ ,\ \ \ \psi =\nu\tau \ ,
}
\eqn\rrtt{
\rho=\rho(\s )=\rho (\s+2\pi )\ ,\ \ \ \ \
\theta=\theta(\s )=\theta (\s+2\pi )\ ,\
}
$$
 \beta={\pi\over 2}\ ,\ \ \ \ \psi_1=\psi_2=\psi_3=\td\phi=0  \ .
$$
It describes a string rotating in $AdS_5$ and in $S^5$ with
independent angular velocity parameters $\om $ and $\nu $, which
is stretched along the radial coordinate and along the angular
coordinate $\theta $ of $S^5$. We will find below that this
rotating soliton generalizes the solutions discussed in
\refs{\gkpsol ,\ft}, and smoothly interpolates beween all previously
known cases.


The equations of motion and constraints become
\eqn\eequa{
\rho '' + (\om^2 -\kappa ^2) \cosh\rho\sinh\rho =0\ ,
}
\eqn\equa{
\theta ''  +\nu ^2 \cos\theta\sin\theta =0\ ,
}
\eqn\coss{
-\kappa^2 \cosh^2\rho + { \rho '}^2+\om^2 \sinh^2\rho + { \theta '}^2
+ \nu ^2 \sin^2\theta =0\ ,
}
where prime denotes derivative with respect to $\s $.
The general solution is given by
\eqn\solu{
{ \rho '}^2=\kappa^2 \cos^2\a_0 -(\om^2-\kappa^2) \sinh^2\rho \ ,
}
\eqn\solv{
 {\theta '}^2
= -\nu ^2 \sin^2\theta +\kappa ^2 \sin^2\a_0\ .
}
where $\a_0 $ is an integration constant.

We have chosen a convenient parametrization in terms of $\cos\a_0 , \sin\a_0 $
to take into account automatically that the solution describes a finite closed string
stretching from $\rho=0 $ up to some $\rho=\rho_{\rm max}$, which is finite provided $\om >\kappa $.
The string is folded onto itself, and the interval $0\leq \s< 2\pi $ is split into four segments.
The first segment starts with $\rho=0 $ at $\s=0$ up to $\rho_{\rm max}$ at $\s={\pi\over 2}$.

Let us define parameters
\eqn\param{
a\equiv {\nu^2 \over \kappa^2 \sin^2\a_0 } \ ,\ \ \ \ b\equiv
{\om^2-\kappa^2\over   \kappa^2\cos^2\a_0 } \ .
}
We will assume $b>0$. There are different situations according to the value of $a$.

The case $a=1$ gives a solution with infinite energy, unless
$\theta \equiv {\pi\over 2}$, which is the case discussed by \ft .
This includes the cases discussed in sects. 2 and 3 of \gkpsol , corresponding to the pp wave limit and to the Regge string.

If $a< 1$, then  $\theta \in [0, \pi] $. This describes a closed string stretched around the great circle of $S^5$. This solution is a generalization of a similar solution (but with
$\rho\equiv  0$) discussed in \gkpsol\ (it reduces to that particular constant $\rho $ case when
 $b\gg 1$). As pointed out in \gkpsol , the solution seems unstable, since a small perturbation
of the string makes it to slide over the sides of the sphere due to the string tension.

 Here we shall consider the interesting case of $a>1$.
Then $\theta \in [0, \theta_{\rm max}] $, where $ \theta_{\rm max}={\rm arcsin}(a^{-1/2}),\ \theta_{\rm max}<{\pi\over 2} $. It generalizes the case discussed in sect.~4.1 of \gkpsol\ (corresponding to the special case $\rho\equiv 0 $, $a>1$)
to strings which also stretch along the radial direction.
In the regime $a\sim 1$, the major contribution to the quantum numbers of the string will come from the region $\theta \sim {\pi\over 2}$. For this reason, in this regime we will recover  the results of \ft\ for the anomalous dimension, with a small correction due to the fact that
the string is also stretched along the $\theta $ direction.

\smallskip

Thus there is a general solution interpolating between the different known cases. This will give a more general formula relating the dimension of the operators with the R-charge and spin.

Note that there are no solutions in this class where
$\theta \in [{\pi\over 2}, {\pi\over 2}+\epsilon ]$. Either $\theta $ is stuck at ${\pi\over 2}$, or
 it goes all the way up to $\theta =0 $.
However, one can consider time-dependent small fluctuations around $\theta={\pi\over 2}$~. This semiclassical quantization was carried out in \gkpsol , \ft , and it leads to the full spectrum \rdos , including the contributions of oscillators.
For the present soliton,
 the semiclassical quantization  in terms of a small $\theta $
expansion, retaining only the harmonic oscillations, is
meaningful only for small $J$, where
$\theta_{\rm max}$ is small,
but not in the interesting region of large $J$ (corresponding to $a\sim 1$), where the angular variable $\theta  $
takes all values from $0 $ to ${\pi\over 2}$.


\medskip

{}From eqs. \solu , \solv , we obtain
\eqn\iner{
(\kappa \sin\a_0)\ \s =\int_0^\theta d\theta' {1\over\sqrt{1-a\sin^2\theta' }}
= {\cal F}(\theta , a)\ ,
}
\eqn\inser{
(\kappa \cos\a_0)\ \s  = \int_0^\rho d\rho' {1\over\sqrt{1-b \sinh^2\rho' }}
= -i {\cal F}(i \rho , -b)\ ,
}
where ${\cal F}(x,m )$ represents the elliptic integral of the first kind. These equations define $\rho =\rho (\theta )$.
In the present case of $a>1$ and $b>0$,  there are  maximum values
of $\theta  $ and $\rho $ taken by the string which are given by
$\theta_{\rm max }={\rm arcsin}(a^{-1/2}),\ \rho _{\rm max}={\rm arcsinh}(b^{-1/2})
$.
{}From now on it is convenient to trade the parameters $\om ,\nu $
by $a,b$ (see \param ).

The periodicity condition \rrtt\ determines $\a_0, \kappa $ in
terms of $a, b$. As a function of $\s $, $\rho (\s ), \ \theta(\s
) $ start at $\rho=0,\ \theta=0$ at $\s=0 $ and reach a maximum
value of $\rho $ and $\theta $ at the point where $\rho ', \theta
'$ vanish, which for the one-fold string is at $\s={\pi\over 2}$.
Therefore one has $\rho (\pi/2 )=\rho_{\rm max}, \ \theta(\pi/2
)= \theta _{\rm max} $. Demanding this,  we get the conditions:
\eqn\peri{
{\pi\over 2}\kappa \sin\a_0 =
\int_0^{\theta_{\rm max}} d\theta {1\over\sqrt{1-a\sin^2\theta }}
={\pi\over 2\sqrt{a} }\ {}_2F_1(\ha,\ha,1;{1\over a})\ ,
}
\eqn\peril{
 {\pi\over 2} \kappa\cos\a_0
=\int_0^{\rho_{\rm max}} d\rho {1\over\sqrt{1-b \sinh^2\rho }}
={\pi\over 2\sqrt{b} }\ {}_2F_1(\ha,\ha,1;-{1\over b})
\ .
}
The hypergeometric function is related to $\K (m ) $,  the complete elliptic integral
of the first kind ($\K (m)={\cal F}(\pi/2 ,m)$).

The energy, spin, and R-charge of the soliton are given by the
following formulas:
\eqn\energia{
E= {R^2\over 2\pi } \kappa \int
_0^{2\pi } d\s \cosh^2\rho =
{2R^2\over \pi \cos\a_0 }
\int
_0^{\rho_{\rm max}}d\rho \ {\cosh^2\rho\over\sqrt{1- b\sinh^2\rho }
} \ ,
}
\eqn\espin{
S= {R^2\over 2\pi }  \int _0^{2\pi } d\s \ \om
\sinh^2\rho =
{2R^2\om \over \pi \kappa \cos\a_0 } \int
_0^{\rho_{\rm max}}
d\rho \ {\sinh^2\rho\over\sqrt{1- b \sinh^2\rho }
} \ ,
}
\eqn\carga{
J={R^2\over 2\pi }  \int _0^{2\pi } d\s \ \nu
\sin^2\theta = {2R^2\nu \over \pi \kappa \sin\a_0 }\int
_0^{\theta_{\rm max}}d\theta \ {\sin^2\theta\over\sqrt{1- a
\sin^2\theta } } \ .
}
Computing the integrals, we find
\eqn\elene{
E=E(a,b)=
{R^2\over  \sqrt{b} \cos\a_0}\ {}_2F_1(-\ha,\ha,1;-{1\over b})\ ,
}
\eqn\esel{
S=S(a,b)=
{R^2\over  2b^{3/2}\cos\a_0}\sqrt{1+b\cos\a_0^2} \
{}_2F_1(\ha,{\textstyle{3\over 2}},2;-{1\over b})\   ,
}
\eqn\edel{ J=J(a)=
{R^2\over 2 a } {}_2F_1(\ha,{\textstyle{3\over 2}},2;{1\over a})\ ,
}
$$
 \tan\a_0 = \sqrt{ {b\over a} }  \  { {}_2F_1(\ha,\ha,1;{1\over a}) \over
  {}_2F_1(\ha,\ha,1;-{1\over b})  } \ .
$$
These formulas define parametrically $E=E(J,S)$.

Let us now derive  explicit analytic formulas for $E=E(J,S)$ in four different regimes, according to the cases $J$ or $S \gg \sqrt{\lambda }$, and $J$ or $S \ll \sqrt{\lambda }$. In all cases, we will consider $J,S \gg 1$.
The regimes of large spin $S$ and large R-charge $J$ (as compared to $\sqrt{\lambda }$) correspond to $b\ll 1$ and to $a\sim 1$ respectively,  whereas
small $S$ and $J$ correspond to $b\gg 1$ and to $a\gg 1$.

Define
$$
I_1(b)=\int _0^{\rho_{\rm max}}d\rho {1\over\sqrt{1- b\sinh^2\rho }}\ ,\ \ \ \ I_2(b)=\int _0^{\rho_{\rm max}}d\rho {\sinh^2\rho\over\sqrt{1- b\sinh^2\rho }}\ ,\ \ \ \
$$
$$
I_3(a)=\int _0^{\theta_{\rm max}}d\theta {1\over\sqrt{1- a\sin^2\theta }}\ ,\ \ \ \
I_4(a)=\int _0^{\theta_{\rm max}}d\theta {\sin^2\theta \over\sqrt{1- a\sin^2\theta }}\ .\ \ \ \
$$
The basic expansion formulas we need are
\eqn\dert{
I_1(b) \cong -\ha \log { b}\ ,\ \ \ \ I_2(b)\cong  {1\over b}+ \four \log
{b},\ \ \ {\rm for} \ \ b\ll 1\ ,
}
\eqn\expa{
I_1(b) \cong {\pi\over 2\sqrt{b}}\big(1-{1\over 4 b}\big) \ ,\ \ \ \
I_2(b)\cong {\pi\over 4b^{3/2} }\ ,\ \ \ {\rm for}\ b\gg 1\ ,
}
and
\eqn\dart{
I_3(a) \cong -\ha \log { {(a-1)\over 16} }\ ,\ \ \ \ I_4(a)\cong  -\ha \log
{{(a-1)\over 16} } -1,\ \ \ \ {\rm for} \ \ a\cong 1\ ,
}
\eqn\dare{
I_3(a) \cong {\pi\over 2\sqrt{a}}\big(1+{1\over 4 a}\big) \ ,\ \ \ \
I_4(a) \cong {\pi\over 4a^{3/2} }\ ,\ \ \ \ {\rm for} \ a\gg 1\ .
}
\medskip

Let us now consider the different cases:
\smallskip

\noindent I) $a\gg 1$, $b\gg 1$, $b/a$ fixed (short strings):
This is the regime of small $S$ and $J$ with fixed $J/S$.
{}From eqs.~\energia, \espin , \carga , we get, to leading order,
\eqn\cortas{
 E\cong R^2\sqrt{a^{-1}+b^{-1}}\ ,\ \ \ \ S\cong {R^2\over 2b}\ ,\ \
\ \ J\cong {R^2\over 2 a}\ ,
}
and $\tan\a_0\cong \sqrt{J/S}$. Thus
\eqn\awa{
E^2 \cong 2\sqrt{\lambda } (J+S) \ ,\ \ \ \ {J\over\sqrt{\lambda }}\ll 1\ ,\ \  {S\over\sqrt{\lambda }}\ll 1\ ,
}
which is the usual Regge-type
spectrum of string theory in flat spacetime. This is expected, since short strings do not feel the curvature of spacetime.
In the case $J/S\to 0$, eq.~\awa\ reduces to the case
discussed in  ref.~\gkpsol .
Equation \awa\ differs from the formula $E^2 \cong J^2+2\sqrt{\lambda }\  S  $
of \ft .
The reason is the following: in the case of  \ft , the string is located at $\theta=\pi/2$; in the present case, the string is located near $\theta =0,\ \rho=0$, where there is a symmetry $J\leftrightarrow S$ due to the symmetry of the solution at small $\rho , \theta $ under $\phi \leftrightarrow \psi  $. From the gauge theory point of view,
the soliton of \ft \ and the present solution correspond to different operators when $J\neq 0$.

\smallskip

\noindent II) $a\sim 1$, $b\ll 1$: This is the case of large R-charge $J$ and large spin $S$. Then the string is  long with $\theta_{\rm max}\sim {\pi\over 2}$.
Now eqs.~\energia \ - \carga \   give
\eqn\poq{
E \cong {2R^2\over\pi \cos\a _0} \bigg({1\over b}- \four \log b \bigg)\ ,
}
\eqn\poqq{
S\cong {2R^2\over\pi \cos\a _0} \bigg({1\over b} + \four \log b \bigg)\ ,
}
\eqn\jjq{
J \cong -{R^2\over \pi }\log {(a-1)\over 16} - {2 R^2\over\pi }\ ,
}
and $\tan\a_0={\log {(a-1)\over 16}\over\log b}$. Thus we obtain
\eqn\smene{
E-S={\sqrt{\lambda }\over\pi } \sqrt{\log ^2 b +  ( {\pi J\over \sqrt{\lambda } } +2)^2
 }\ .
}
The parameter $b$ is a function of $S$ and $J$, determined by
the transcendental equation
\eqn\tras{
S={2 \sqrt{\lambda }\over \pi b|\log b |} \sqrt{ \log ^2 b +  ( {\pi J\over \sqrt{\lambda } } +2)^2 } \ .
}
For $ {J\over \sqrt{\lambda }}  \ll \log {S\over\sqrt{\lambda } } $,  we have $b\cong  {2 \sqrt{\lambda }\over \pi S}$.
Thus we obtain
$$
E-S \cong  \sqrt{ (J +  {2\sqrt{\lambda }\over \pi } )^2 +{\lambda \over \pi ^2}
\log^2 {\pi S\over 2\sqrt{\lambda} } }\
$$
\eqn\espec{
\cong {\sqrt{ \lambda }\over \pi } \log {\pi S\over 2\sqrt{\lambda} } +
{\pi (J+2\sqrt{\lambda }/\pi )^2\over 2\sqrt{\lambda } \log {\pi S\over 2\sqrt{\lambda} } }\ ,
}
$$
{J\over\sqrt{\lambda }}\gg 1\ ,\ \ \ {S\over\sqrt{\lambda }}\gg 1\ ,\ \ \
{ J\over \sqrt{\lambda }}  \ll \log {S\over\sqrt{\lambda } } \ .
$$
Corrections to this formula are of order $J^2/\log^2 {S\over\sqrt{\lambda} }$.

For $ { J\over \sqrt{\lambda }}  \gg \log {S\over\sqrt{\lambda } } $,
we have $-b\log b={2J\over S}$, so that
\eqn\otra{
E-S \cong J +{2\sqrt{\lambda }\over \pi } + {\lambda \over 2\pi ^2
J } \log^2 {2J\over S}\ ,
}
$$
{J\over\sqrt{\lambda }}\gg 1\ ,\ \ \ {S\over\sqrt{\lambda }}\gg 1\ ,\ \ \
{ J\over \sqrt{\lambda }}  \gg \log {S\over\sqrt{\lambda } } \ .
$$

The formulas \espec , \otra\ agree with the corresponding formulas in \ft\ for a rotating string fixed at $\theta={\pi \over 2}$.
The reason is that in this regime $a\sim 1$, most of the contribution to the R-charge comes from the region $\theta ={\pi \over 2}$.
The only difference is the shift in $J$ by ${2\sqrt{\lambda }\over \pi }$, which represents the subleading correction to the large $J/\sqrt{\lambda }$ expansion.
As mentioned above, the presence of this term reflects the fact that
the string state we are considering is not the same as the state considered in \ft .


\smallskip

\noindent III) $b\ll 1, \ a\gg 1 $: In this case  ${J\over\sqrt{\lambda }}\ll 1\ ,\  \
{S\over\sqrt{\lambda }}\gg 1 $;  the string is
long with $\theta$ being nearly fixed at $\theta \cong 0$. Now $\tan\a_0={\pi \over \sqrt{a}\ |\log b|}$.
Using eqs.~\poq , \poqq,
and $J$ as in \cortas , we obtain
\eqn\zmene{
E-S= {\sqrt{\lambda }\over \pi }|\log b |(1 + {\pi ^2J\over
\sqrt{\lambda }\log^2 b} ) \ ,
}
with $b\cong  {2 \sqrt{\lambda }\over \pi S}$.
 Thus
\eqn\jjo{
E-S={\sqrt{\lambda } \over \pi } \log {S\over\sqrt{\lambda} }+{\pi J\over \log {\pi S\over 2\sqrt{\lambda } } }  \
, }
$$
{J\over\sqrt{\lambda }}\ll 1\ ,\ \ \ {S\over\sqrt{\lambda }}\gg 1\ .
$$
This formula is new, and differs from the analog limit (small $J$, large $S$) in \ft . Again, the reason is that the small $J$ string  spins in the region of small $\theta $, whereas in \ft\ is always fixed at $\theta ={\pi \over 2} $.

\smallskip

\noindent IV) $b\gg 1,\ a\sim 1 $. Then the string extends from
$\theta=0 $ to $\theta_{\rm max}\sim {\pi\over 2}$ with $\rho $
being nearly fixed at $\rho \cong 0$. Now $ {J\over\sqrt{\lambda
}}\gg 1 $ and  $ {S\over\sqrt{\lambda }}\ll 1 $. Using
eqs.~\energia , \espin , \carga , we obtain
\eqn\enf{
E= {2\sqrt{\lambda
}\over \pi } I_3 \sqrt{1+{I_1^2\over I_3^2} }\ (1+{I_2\over I_1}) \ ,
}
\eqn\esf{
S= {2\sqrt{\lambda }\over \pi } I_3 \ {I_2\over
I_1}\sqrt{1+ (1+b){I_1^2\over I_3^2} } \ ,
}
\eqn\jjf{ J\cong
{2\sqrt{\lambda }\over \pi } (I_3-1) \ .
}
 Combining these equations, we
find
\eqn\ffg{
E\cong J + {2\sqrt{\lambda }\over \pi } + S+{\lambda  S\over 2 J^2 } \ ,
}
$$
{J\over\sqrt{\lambda }}\gg 1 \ ,\ \ \ \  {S\over\sqrt{\lambda }}\ll 1\ .
$$
For  $S=0$, eq.~\ffg\ reduces to the result \rtres \ of \gkpsol .
Equation \ffg\ can be compared with the similar formula found in \ft  ,
\eqn\zzg{
E\cong J +  S+{\lambda  S\over 2 J^2 } \ .
}
Note that in the present case the correction
${2\sqrt{\lambda }\over \pi }$ is important, being larger than the next terms.

\medskip

\noindent V) General case: Using the general formulas
eqs.~\elene \ - \edel , one can do a
numerical plot of $E=E(S,J)$ in the general case. We find that  $E=E(S,J)$
smoothly interpolates between  the different regimes described above,
with no surprises at intermediates regimes.

\newsec{Discussion}

The regime  IV) is closely related to strings in the
 pp-wave background. Indeed, this background is obtained as the limiting geometry seen by a particle
 moving along the $\psi $ direction with large momentum $J$, sitting near $\rho =0 $ and near
 $\theta  ={\pi\over 2}$ \refs{\blap,\mald }.\foot{
Note that our definition of $\theta $ in \metric\ differs by
 $\theta\to {\pi\over 2}-\theta $ from the notation of \mald .
} To see that \zzg \ can be understood from the string spectrum in
the pp wave background, we start with eq.~\rdos , and consider
string states with spin $S$ associated with rotations in the plane
1-2. Let us assume that it is a state of the Regge trajectory, so
that only $n=1$ oscillators in the directions 1-2 are excited
(e.g. by acting $S/2$ times with $a_{1+}^\dagger , \ \tilde
a_{1+}^\dagger , $ on the light-cone vacuum, where $a_{n+}^\dagger
={1\over\sqrt{2}}(a_n^{1\dagger}- i a_n^{2\dagger})$~). Then we
have $N_n= S+ N^T_n$, where $N^T_n$ contains the oscillators
corresponding to the other directions. Thus eq.~\rdos \ becomes
\eqn\zdos{ E=J + S\sqrt{1+{\lambda \over J^2} } +
\sum_{n=-\infty}^\infty  N_n^T \sqrt{1+{\lambda n^2\over J^2} }\ ,
} For $J\gg \sqrt{\lambda }$, this gives \eqn\hhpp{ E\cong J +S +
{\lambda S\over 2J^2} + \sum_{n=-\infty}^\infty N_n^T
\sqrt{1+{\lambda n^2\over J^2} }\ . } This agrees with eq. \zzg ,
up to the contribution of extra oscillators. This quantum part of
the spectrum can be captured by a
 semiclassical quantization around the soliton \refs{\gkpsol,\ft }.

Given the correspondence between physical string states and gauge
theory operators, there must be operators in the dual gauge theory
with dimension given by \eqn\hhp{ \Delta \cong J +S + {\lambda
S\over 2J^2}+... \  . } Denoting by $\phi^1,...,\phi ^6$ the six
scalars of $\N= 4$ Yang-Mills theory and $Z=\phi^5+i \phi ^6$, the
unique  single trace operator of bare dimension $\Delta=J $ is
given by  \mald\ ${\rm tr} [Z^J]$.
 The spin $S$ can be introduced by adding covariant derivatives
to this  operator, i.e. replacing $S$ factors $Z$ by  $D_i Z=\p_i Z+[A_i,Z]$,
$i=1,2 $,
 adding phase factors of the form $\exp (2\pi i l/J)$, and summing
 over all possible insertion points ($l$ is the position of $D_i$ along the string of $Z$'s).
We propose that the resulting operator: \eqn\jop{ {\cal
O}_{i_1...i_S}= \sum_{l_1,...,l_s=1}^J {1\over \sqrt{J} N^{J/2} }
\ {\rm tr}\big[...ZD_{i_1}Z...ZD_{i_S}Z...\big] e^{ {2\pi i\over
J}  (l_1+...+l_S) } \ ,\ \ \ \ S\ll  J\ , }
 should be identified with the soliton of \ft \
(which is fixed at $\theta ={\pi\over 2}$). One of the sums can be
performed by using the cyclic property of the trace.
 For $S\ll J$, this operator is almost BPS. The bare dimension is
$\Delta=J+S$. The correction ${\lambda S\over 2J^2}$  arises by a
one-loop calculation,
 by considering  a subset of Feynman diagrams, as follows.
Each insertion  of $D_i$ (corresponding to  $a^\dagger_{1+}$ acting on the
light-cone vacuum)
 is treated similarly as the
 insertions of $\phi ^r$ operators computed in \mald .\foot{
In the pp wave background there is a symmetry under the exchange
of the directions 1234 and 5678, which in the gauge theory
corresponds to the
exchanges of $D_i$ with $\phi ^r$.} They give  the contribution
to the anomalous dimension
$$
(\Delta-J)_1=1+ {\lambda \over 2J^2}\ ,
$$
so that, for $S$ insertions, we get $\Delta-J=S+{\lambda S\over
2J^2}$, in agreement with eq.~\hhp . This gives evidence that
these operators are not decoupled in the large $\lambda $ limit,
with $\lambda/J^2 $ fixed.

Let us now consider
the string of section 2, which is stretched from
$\theta=0 $ up to $\theta = \theta_{\rm max}$.
Consider the case $S=0$ and $J\gg \sqrt{\lambda }$.
The correction to the energy
by a shift ${2\sqrt{\lambda }\over \pi }$ in \ffg\ has a simple
interpretation.
For a string of large
$J$ which is not stretched in the $\theta  $ direction, its energy
is $E=J$. When this string is  stretched from $\theta=0 $ up to
$\theta ={\pi\over 2}$, its energy must increase in a quantity
approximately given by $E\cong J+ {\rm tension}\times {\rm length}$.
Indeed, using \iner , the energy \energia\ can also be written as
\eqn\nuen{ E= {2\sqrt{\lambda }\over \pi \sin\a_0} \int
_0^{\theta_{\rm max} }d\theta\ {1\over \sqrt{1-a \sin^2\theta }
}\cong J+ {\sqrt{\lambda }\over 2\pi }\ 4\int _0^{ {\pi\over 2}
}d\theta\ \cos\theta  \ ,
} where
we have used \carga\ and $a\sim 1$. Thus we recover eq.~\ffg\ for
$S=0$, $E\cong J+{2\sqrt{\lambda }\over \pi }$.


The possible decay of the string of section 2 into BPS states should be suppressed by powers of the string coupling $g_s$. {}From the gauge theory point of view, such string
should correspond to a highly excited
local operator. This operator should be a linear combination of
operators containing  $S$ insertions of $D_i$
 into ${\rm tr}[Z^J]$
--~to account for the spin as above~-- and, in addition,
insertions of various $\phi ^r $, $r=1,2,3,4$. It remains an
interesting question to identify the corresponding operator.\foot{
 String states with $E-J\sim
\sqrt{\lambda }$ can be explicitly constructed in the pp-wave
background in terms of coherent states of length proportional to
$R=\lambda^{1/4}$.}


\bigskip

\noindent {\bf Acknowledgements}

\noindent The author wishes to thank J.~Gomis and A.~Tseytlin for useful remarks.

\vfill\eject \listrefs
\end